\documentclass{aa}  

\usepackage{graphicx}
\usepackage{txfonts}
\usepackage{hyperref}

\makeatletter
\renewcommand*\aa@pageof{, page \thepage{} of \pageref*{LastPage}}
\makeatother

\newcommand{\bpb}{$\beta$ Pictoris b}

\begin{document}

   \title{bRing: An observatory dedicated to monitoring\\ the $\beta$ Pictoris b Hill sphere transit}


   \author{R. Stuik
          \inst{1}
          \and
          J.I. Bailey, III
          \inst{1}
          \and
          P. Dorval
          \inst{1}
          \and
          G.J.J. Talens
          \inst{1}
          \and
          I. Laginja
          \inst{1}
          \and
          S. N. Mellon
          \inst{2}
          \and
          B.B.D. Lomberg
          \inst{3,4}
          \and
          S.M. Crawford
          \inst{3}
          \and
          M.J. Ireland
          \inst{6}
          \and 
          E.E. Mamajek
          \inst{5,2}
          \and
          M.A. Kenworthy
          \inst{1}
          }

   \institute{Leiden Observatory, Leiden University, P.O. Box 9513, 2300 RA Leiden, The Netherlands\\
              \email{stuik@strw.leidenuniv.nl}
         \and    
             Department of Physics \& Astronomy, University of Rochester, 500 Wilson Blvd., Rochester, NY 14627-0171, USA
         \and
             South African Astronomical Observatory, Observatory Rd, Observatory Cape Town, South Africa
         \and    
             University of Cape Town, Rondebosch, Cape Town, 7700, South Africa
         \and    
             Jet Propulsion Laboratory, California Institute of Technology, M/S 321-100, 4800 Oak Grove Drive, Pasadena, CA 91109, USA
         \and 
         	Research School of Astronomy and Astrophysics, Australian National University, Canberra, ACT 2611, Australia\\
             }

   \date{Received 2017 July 30; accepted 2017 August 28}

\titlerunning{bRing}

 
  \abstract
   {}
   {We describe the design and first light observations from the $\beta$
Pictoris b Ring (``bRing'') project. The primary goal is to detect
photometric variability from the young star $\beta$ Pictoris due to
circumplanetary material surrounding the directly imaged young extrasolar gas giant planet \bpb.}
   {Over a nine month period centred on September 2017, the Hill sphere of the planet will cross in front of the star, providing a unique opportunity to directly probe the circumplanetary environment of a directly imaged planet through photometric and spectroscopic variations. 
We have built and installed the first of two bRing monitoring stations
(one in South Africa and the other in Australia) that will measure the flux of $\beta$ Pictoris, with a photometric precision of $0.5\%$ over 5 minutes.
Each station uses two wide field cameras to cover the declination of the star at all elevations.
Detection of photometric fluctuations will trigger spectroscopic observations with large aperture telescopes in order to determine the gas and dust composition in a system at the end of the planet-forming era.}
   {The first three months of operation demonstrate that bRing can obtain better than 0.5\% photometry on $\beta$ Pictoris in five minutes and is sensitive to nightly trends enabling the detection of any transiting material within the Hill sphere of the exoplanet.}
   {}


   \maketitle
%

\section{Introduction}

Planet formation theory predicts that gas giant planets form in
circumstellar disks which disperse on the timescale $\sim10^7$ Myr
\citep[see e.g. ][]{Mamajek09,Bell13,Pecaut16} and that these planets
have circumplanetary moon-forming disks around them
\citep{Canup02,Alibert05,Heller15b,Heller15,Heller15c}. 
Studying young stars provide an opportunity to capture planetary systems in formation, and study how circumplanetary disks collapse into moons orbiting these exoplanets.
A complex eclipse seen towards the young star 1SWASP
J140747.93-394542.6, also now designated V1400 Cen
\citep{Mamajek12}, is hypothesised to be a giant ring system filling most
of the Hill sphere of an as yet undetected substellar secondary companion \citep{Kenworthy15}.
This eclipse may be the first detection of a circumplanetary disk undergoing dynamical sculpting by forming satellites, and these giant ring systems provide a snapshot of a circumplanetary environment at a very early epoch.
The young star PDS 110 was recently discovered to undergo regular eclipses that could be due to an extended disk like structure around a secondary companion \citep{2017arXiv170510346O}.
Exomoons have been extensively searched for in the Kepler satellite data by \citet{Kipping12}, with a tentative detection of an exomoon transit reported in \citet{Teachey17}.
Searching for more Hill-sphere transits of young planets provides an opportunity for transmission spectroscopy of the gas and dust in a circumplanetary environment.

The young \citep[$\sim23$\,Myr old;][]{Mamajek14} star $\beta$ Pictoris
hosts a near edge-on dusty debris disk \citep[first directly imaged by][]{Smith84} and a 13 $M_{Jup}$ planet \citep{Lagrange09,Lagrange09-1, Chilcote17}.
The planet has been measured to have an orbital period of approximately 20 years through astrometric monitoring \citep{chauvin12, nielsen14, Millarblanchaer15}, and the Hill sphere of the planet will transit from April 2017 to March 2018 \citep{Wang16}, lasting approximately 300 days, with an impact parameter of 20\% of the Hill radius.
This presents a unique opportunity to probe the circumplanetary environment of a young giant exoplanet with photometry and spectroscopy.

In 1981 the star underwent a gradual brightening and fading over a few
weeks, centered on a rapid photometric variability event, of around 6\%
\citep{Lamers97,LecavelierdesEtangs97}.
The two main hypotheses for the event were a planetary transit or scattering by a dust cloud \citep{Lamers97}.
While  extended structures have also been seen in transit around other
stars \citep[.e.g.][]{Graczyk03,Galan10,Chiang04,Winn04, Rodriguez15,
Rappaport12,Brogi12}, the upcoming inferior conjuction of $\beta$
Pictoris b is an excellent opportunity to test the hypothesis that a
circumplanetery disk was the cause of the 1981 event
\citep{lecavelierdesetangs09,Etangs16}.
Confirmation of circumplanetary material in the Hill sphere of $\beta$ Pictoris b would provide evidence for the third such system around a planet orbiting a young star. 

We have designed a novel transit experiment called the {\it $\beta$ Pictoris b Ring} project, (shortened to ``{\it bRing}'') consisting of two observatories photometrically monitoring the young, nearby star $\beta$ Pictoris during the transit of the Hill sphere of its giant exoplanet 
$\beta$ Pictoris b in 2017-2018. 
This paper describes the commissioning and installation of the first bRing observatory, located in the Sutherland observing station of the South African Astronomical Observatory.
In Section \ref{infconj} we summarize the upcoming Hill sphere transit and timescales involved for transit times for  circumplanetary material.
We present the eclipse profiles for plausible disk and ring geometries in Section \ref{ring_models} where lanes are cleared out of the circumplanetary material by the gravitational scattering of exomoons.
In Section \ref{bring_instrument}, we describe the bRing station setup and the location of the South African site.
In Section \ref{firstlight}, the first light curves from January and February 2017 are presented, showing that we reach the photometric precision required for triggering of spectroscopic observations on larger telescopes, and our conclusions in Section \ref{conclusions}.


\section{\texorpdfstring{$\beta$}{B} Pictoris b and the inferior conjunction of 2017}\label{infconj}

{\bf The star:} $\beta$ Pictoris is a bright \citep[V\,=\,3.85;][]{Crifo97}, southern
($\alpha = 05h47m17.08769s$  $\delta=-51^{\circ}03' 59.4412"$), nearby \citep[$d$\,=\,$19.44\,\pm\,0.05$ pc;][]{vanLeeuwen07b}, young \citep[23\,$\pm$\,3 Myr;][]{Mamajek14, 2016MNRAS.455.3345B}, hot star 
\citep[T$_{\rm eff}$\,=\,8052\,K; A6V;][]{Gray06}.  The star has been reported to have $\delta$ Scuti like pulsations \citep{2003MNRAS.344.1250K}.

\noindent{\bf Circumstellar Environment:}
In 1983, the IRAS satellite detected a strong infra-red excess above that expected for the star's photosphere \citep{Aumann84}.
Ground-based coronagraphic imaging observations by \citet{Smith84} showed the presence of an extended disk of dust and gas surrounding the star out to a distance of 500\,au, seen edge-on to our line of sight.  
%
%
Subsequent observations  show a warp in the disk at $\sim50$ au, implying the presence of a massive companion \citep{1995AAS...187.3205B, 1995AJ....110..794K, 1997MNRAS.292..896M, 2000ApJ...539..435H, 2006AJ....131.3109G}.
Circumstellar material is further detected through the presence of
infalling comets \citep{2000Icar..143..170B, 2014Natur.514..462K} and
gas phase detections of the disk \citep{Hobbs85,Vidal-Madjar86,lecavelierdesetangs00,2001Natur.412..706L,2017A&A...599A..75W}.
A recent study can be found in \citet{Apai15}.

\noindent{\bf The planet:} A gas giant planet was detected with direct imaging \citep{Lagrange09,Lagrange09-1} with a location close to the mid-plane of the debris disk.
Subsequent direct and coronagraphic imaging of the planet \citep{Currie13,Morzinski15} shows that the planet's projected separation is decreasing, with the orbit almost edge on to our line of sight.
Radial velocity measurements by \citet{Snellen14} show that the planet's projected velocity is blue shifted, and that the planet is moving towards inferior conjunction.
Astrometric monitoring indicates a period of approximately 20 years for the planet's orbit \citep{chauvin12, nielsen14, Millarblanchaer15}.
Recent spectroscopy between 1 to 2.5 micron of the planet \citep{Chilcote17} assigns it a spectral type of L2 and a mass of about 13 $M_{Jup}$, and rotational broadening of absorption lines in the planets atmosphere show that it is a rapid rotator\citep{Snellen14}.

\noindent{\bf Inferior conjunction:} An astrometric orbital analysis by \citet{Wang16} shows that the planet will not transit the star (at $>$10$\sigma$), but that the star will move behind the planet's Hill sphere with an impact parameter of approximately 20\% of the Hill sphere's radius, enabling the opportunity for transit photometry and spectroscopy of any circumplanetary material that lies between the Earth and stellar disk.
The Hill sphere will begin transit at the start of April in 2017 through to January 2018. 
The transit of $\beta$ Pictoris b's Hill sphere is a unique opportunity for investigating the circumplanetary material of a young, evolving exoplanet.

The hypothesis of Hill sphere material is strengthened by the detection of photometric fluctuations towards $\beta$ Pictoris in 1981 \citep{LecavelierdesEtangs95} which were interpreted as a planetary transit \citep{LecavelierdesEtangs97} or a dust cloud \citep{Lamers97}.
The fluctuations were on the order of 6\% variability over a period of 4 days, superimposed on a gradual brightening and fading over three weeks.


%

\section{Photometric variability due to Hill Sphere filling rings}\label{ring_models}

In order to model the photometric signal we might expect from the Hill sphere transit of $\beta$ Pictoris b, we simulated light curves from the transit of a giant ring system around the planet with a non-zero obliquity, analogous to the model of J1407b \citep{Kenworthy15,Kenworthy15b}.
Due to $\beta$ Pictoris b's young age we assume that it is plausible that a disk of circumplanetary debris exists around the planet out to a large fraction of the Hill radius.
Furthermore, we assume that the rings are coplanar with the Laplace invariable plane of the planet (usually the equatorial plane) out to a significant fraction of the Hill sphere.
The coplanarity of the rings breaks down at a radius from the planet where the torque from the star overcomes the J2 component of the planet's spheroid \citep[e.g. equation 15 of ][]{Dobrovolskis93}.
For $J_2=0.02$ (slightly more extreme than Jupiter or Saturn), $R=2R_J$, $a= 9.2au$, $M_P/M_*=0.004$, we get a critical radius of 0.04 AU - about 4\% of the Hill radius.
A more recent analysis presented in \citet{Zanazzi17} shows that this radius can be larger if there is significant self-gravity within the disk, but with $\beta$ Pictoris b is is not clear that this is the case.
Material outside this radius in a ring or circumplanetary disk aligns with the circumstellar disk, not the planet. 
It is difficult to assess the probability of a high obliquity in a young giant exoplanet.
For solar system planets, there have been suggestions that obliquities can be driven by secular interactions with other planets \citep{Hamilton04}, with low inclinations in the first $\sim 100$ Myr.
Such a hypothesis in our solar system does not, however, explain the J1407 system.
If there are moons orbiting the planet, these satellites may carve out gaps in the disk along their orbits, produce ring-like structures which would then be detectable as they transit the disk of the star.

Simulations of moon formation within circumplanetary disks predict that in 80\% of the time, gas giants will most likely have four to six moons\citep{Heller14} and that the masses of these moons are likely to scale with a factor of $10^{-5}$ to $10^{-4}$ with their host planet mass \citep{Canup06}.
We therefore use the three inner Galilean moons as templates for the exomoons we use in our simulations.
The masses of Io, Europa and Ganymede have been multiplied by 13 to match the mass of beta Pictoris b, and their semi-major axes scaled up to an extent that they still lie in the inner 50\% of $\beta$ Pictoris b's Hill sphere.
We then assume these satellites will clear out gaps within this dust disk consistent with the diameter of their Hill spheres.
The Hill sphere radius for the moons scales with their orbital semi-major axis and with their mass, resulting in a strong detection bias for massive moons at large distances from their parent planet.
See Table \ref{tab:satellites} for the satellite parameters we used in the simulations.
Intermediate dust lanes, filling the inner third of each gap have then been added, as dynamic simulations have shown that when moons clear out gaps in this fashion, there is an agglomeration of dust in the centre of these gaps following horseshoe orbits (Kenworthy et al., in prep.).
The intermediate dust lane, convolved with the finite diameter of the star, results in a characteristic ``double peak'' in the light curve.

\begin{table}
\centering
\caption{Satellites used in the double gap simulation. Mass [1] is given in units of $\beta$ Pictoris b mass, mass [2] in units of Earth masses, the semi-major axis a in percent of the planet's Hill radius and the moons' full Hill sphere transit times in hours.}
\label{tab:satellites}
\begin{tabular}{lcccc}\hline \hline
\textbf{$\beta$ Pictoris b}  & Mass [1] & Mass [2] & $a$ & $t_{Hill}$ \\\hline
	Satellite 1  &  $4.28\times10^{-5}$  &  0.15  &  12\%  &  24\\
	Satellite 2  &  $2.30\times10^{-5}$  &  0.08  &  20\%  &  31\\
	Satellite 3  &  $7.09\times10^{-5}$  &  0.25  &  32\%  &  71\\\hline
\end{tabular}
\end{table}

\begin{figure}
	\centering
	\includegraphics[scale=0.4]{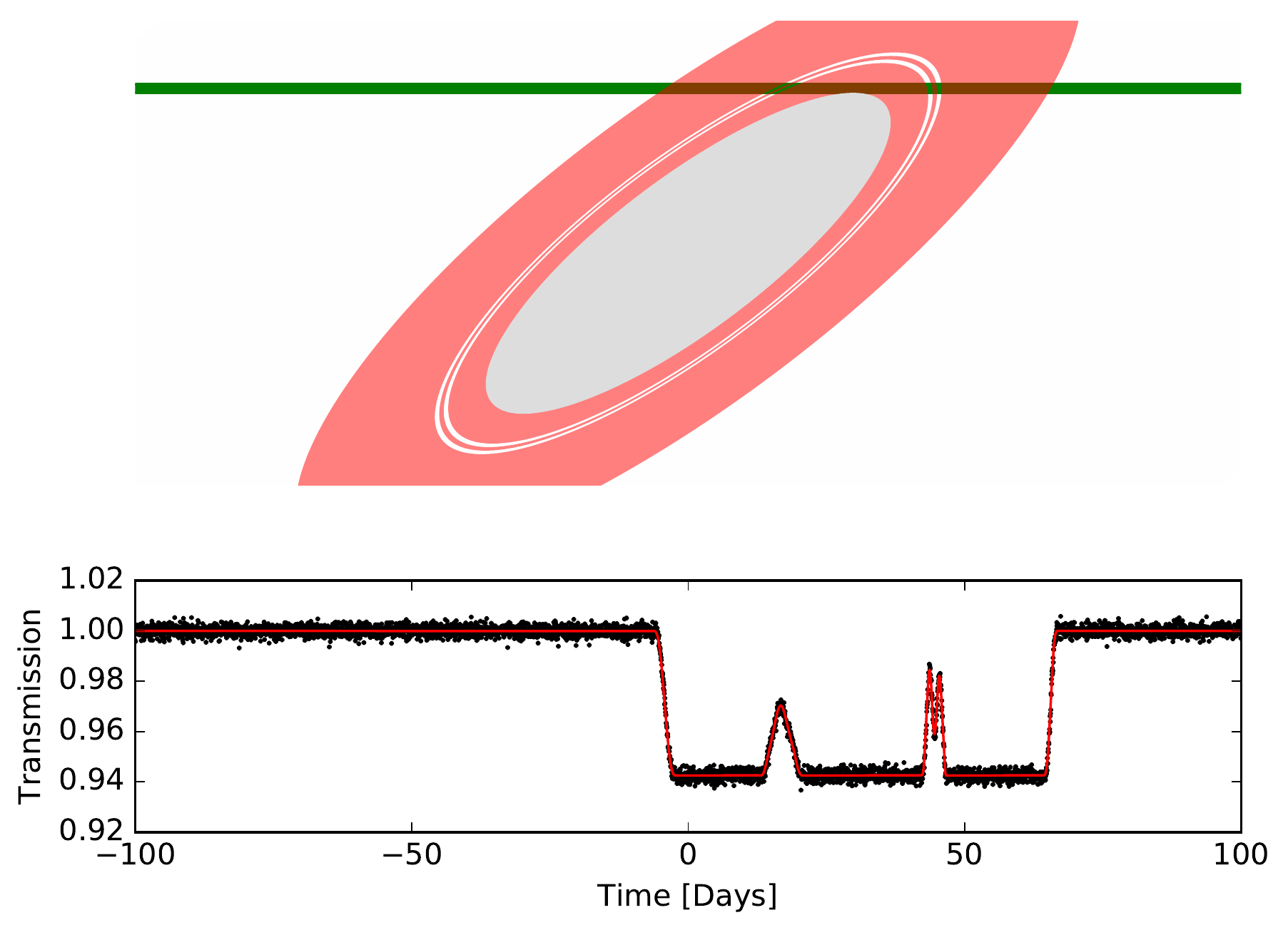}
	\caption{Ring system simulation for the satellites from Table \ref{tab:satellites}. $v_{orb} = 13.3 km/s$, $b = 17\%$ of the Hill radius, which is filled with dust out to 50\% of the Hill radius. 
Inclination $i = 18^{\circ}$ and tilt is $37^{\circ}$ The red rings are filled with dust blocking 6\% of incident flux, white rings are gaps cleared of dust by the moons. The grey zone in the middle is not captured by the transit. The green stripe is the path of the star behind the ring system.}
	\label{ring_system}
\end{figure}

The simulations were computed for an orbital velocity of the planet of 13.3 $km.s^{-1}$ (Jason Wang, private communication).
The orbital velocity of the planet means that any circumplanetary material will take approximately two days to transit the stellar disk.
The impact parameter $b$ of the system and hence the projected distance between the star behind the ring system and its centre is 10 milliarcseconds.
The obliquity of the ring system needs to be at least $18^\circ$ for any circumplanetary dust at the edge of the Hill sphere to transit the star, and higher obliquities will cause transits of rings at smaller radii down to about 20\% of the Hill sphere for face-on rings.
We justify our assumption of non-zero obliquities of the rings based on the wide range of obliquities seen in our Solar system.
A smooth screen of absorption of 6\% is assumed, consistent with the change in brightness seen in 1981 by \citet{LecavelierdesEtangs95}, and the gaps cleared out by the satellites are assumed to be fully free of dust, except for the previously mentioned central part of the gaps.
After creating the ring system, it was convolved with the star on its path behind it and the rings and the resulting light curve can be seen in Figure \ref{ring_system}.
No forward scattering is used in this toy model.
Gaussian distributed noise is added to simulate a photometric precision of 0.5\% every 5 minutes, consistent with the measured sensitivity of bRing.

\subsection*{1981 event geometry}

\begin{figure}
	\centering
	\includegraphics[scale=0.4]{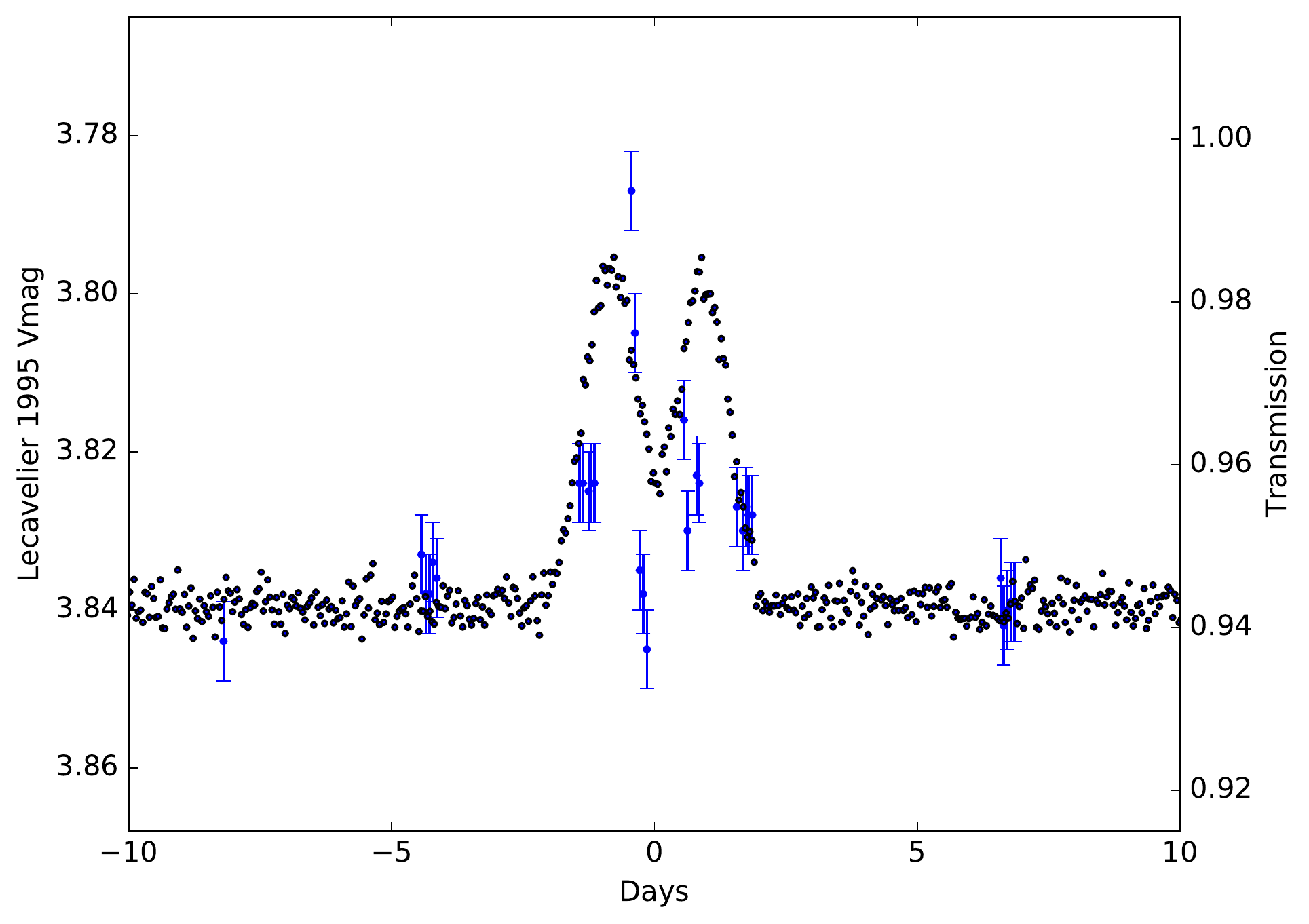}
	\caption{Double gap lightcuvre feature from Fig. \ref{ring_system} ($i = 18^{\circ}$ and tilt = $37^{\circ}$) compared to the November 1981 data, centered on November $10^{th}$ 1981, reproduced from \citep{LecavelierdesEtangs95}.}
	\label{moon_Lecav1}
\end{figure}

Considering the geometry of the ring system described above, the detectability of the satellite gaps is highly influenced by the system's inclination angle and tilt.
Astrometric measurements from \cite{Wang16} led to the conclusion that a half filled Hill sphere would graze the star if a minimum inclination angle of $18^{\circ}$ is given.
If furthermore this configuration is tilted by an angle of $37^{\circ}$ from the direction of projected orbital motion either way, the star has a perpendicular ring edge crossing 45 days before or after the time of minimum approach - see Figure \ref{ring_system}. 
The light curve features emerging from a crossing like that show similarities with the photometric fluctuations in November 1981 \citep{LecavelierdesEtangs95} and could explain them with a double-ring feature transit due to an exomoon, as seen in Figure \ref{moon_Lecav1}.

This ring configuration however is not the only solution that fits the lightcurve shape of the 1981 data.
For lower inclinations like described above, the tilt angle which allows for a fit is constrained.
If higher inclinations are present, the range of tilts that create the suggested feature is larger.
Double gap features from different ring geometries differ in the width of the feature and the depth of the central light curve dip, which can help to constrain the possible ring configurations.

\section{bRing}\label{bring_instrument}

The design of bRing is based on the earlier on-sky validated design of MASCARA \citep{snellen12,snellen13,stuik14,stuik16,Talens17}, which uses five cameras to take images of the whole visible sky above airmass 2 and to search for transiting exoplanets around bright stars $(4 < m_v < 8)$. 
The primary science goal of bRing is to monitor $\beta$ Pictoris during the Hill sphere transit of $\beta$ Pictoris b and to issue triggers for detailed observations if a transiting event is detected. 
The photometric precision goal of $\beta$ Pictoris is 0.5\% every 5 minutes.
$\beta$ Pictoris is located at a declination of -51 degrees and as such requires only two cameras with the field of view of the cameras used for MASCARA.
Within a $\approx$ 53 degrees wide band centered on $\beta$ Pictoris, bRing monitors an additional 20,000 bright stars and performs transmission and cloud monitoring over its field of view.
To get longitudinal coverage, two separate bRing stations are being built (one by Leiden Observatory for the Northern Cape, Sutherland, South Africa location and other by University of Rochester for monitoring in Australia) and the data will be combined with MASCARA South to both provide full, 24-hour coverage, and redundancy.

\subsection{Optics}

\begin{figure}
	\centering
	\includegraphics[scale=0.4]{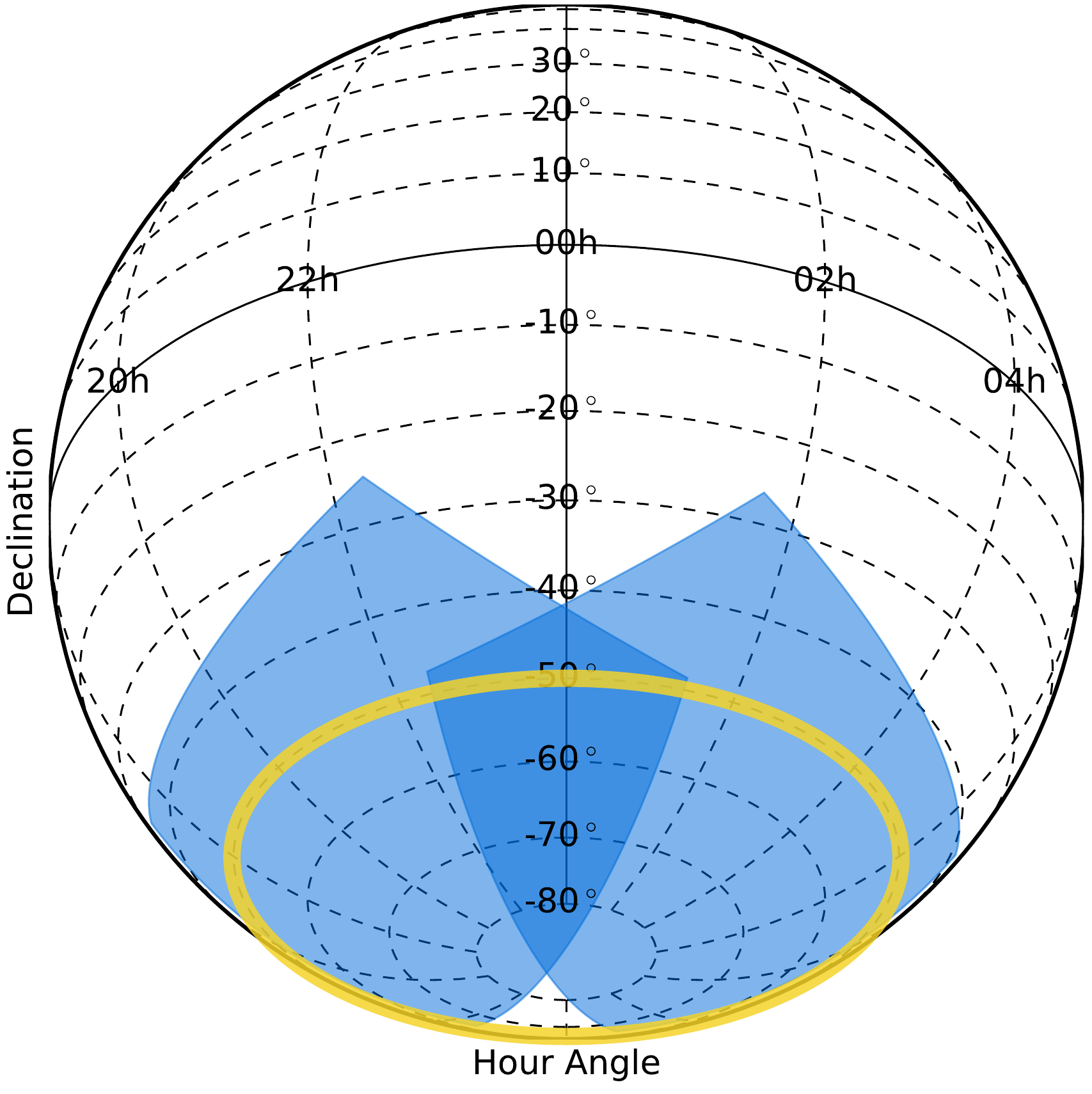}
	\caption{Fields of view of the two bRing cameras on the celestial sphere as seen from South Africa. The fields of view are shown by the blue rectangles, and the yellow line marks the declination of $\beta$ Pictoris.}
	\label{bring_fov}
\end{figure}

The design of bRing composes of two FLI Microline ML11002M Grade 2 Monochrome CCD (Charged Coupled Device) cameras.
These CCDs are equipped with a 16-bit chip of size 24 by 36 mm, with $4008 \times 2672$ square pixels of $9\mu m$ in size.
The cameras are each fitted with a $24mm$ $f/1.4$ wide-field Canon lens \citep{snellen13,stuik14}.
The bRing station faces due South with the two cameras pointing at an elevation of 45 degrees and azimuths of 150 and 210 degrees respectively.
Their fields of view are 74 degrees by 53 degrees and are oriented such that they cover the declination of $\beta$ Pictoris down to an airmass of 4 (see Figure \ref{bring_fov}).

Each camera lens looks through its own glass window, providing environmental isolation.
The windows are made from broadband anti-reflection coated float glass with a thickness of 3 mm and diameter 127 mm  with a surface flatness of $\lambda/4$ at 500nm.

\subsection{Observing Cadence}

\begin{figure}
	\centering
	\includegraphics[scale=0.5]{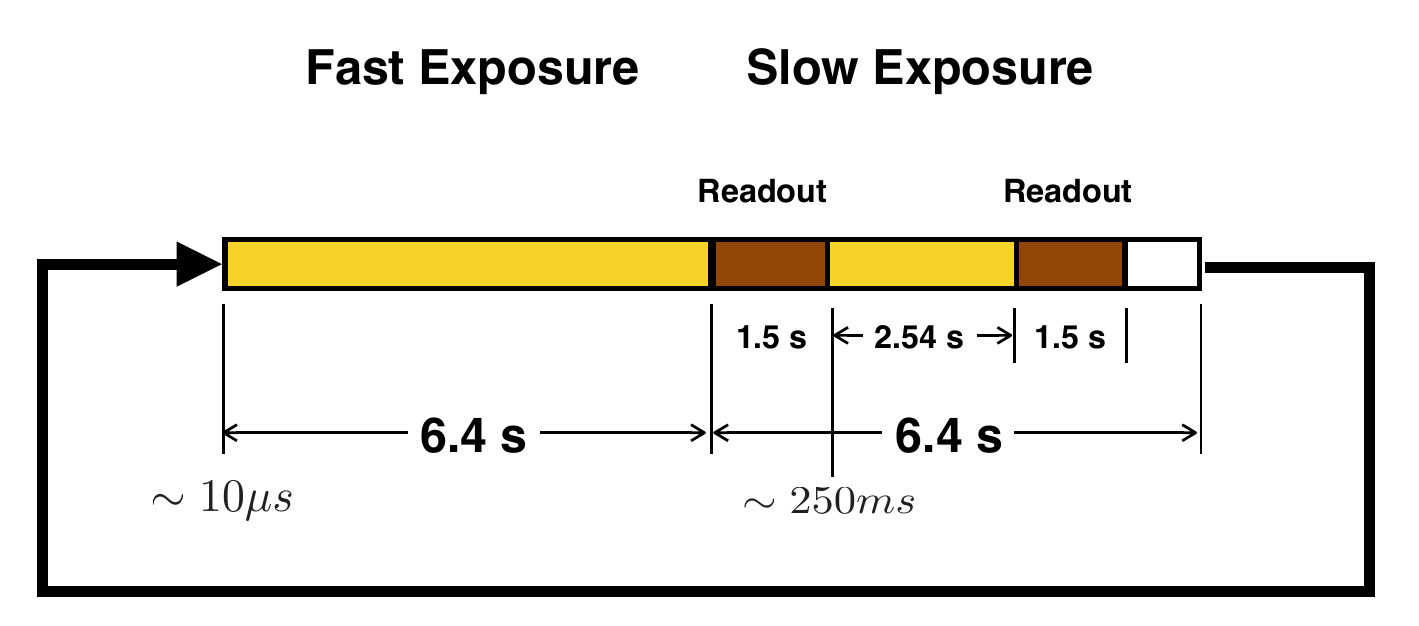}
	\caption{Timing diagram of bRing showing the observing cadence for the fast and slow exposure times. Note: seconds in figure are sidereal seconds.}
	\label{fig:timing}
\end{figure}

In order to cover the dynamic range in brightness from $\beta$ Pictoris (V=3.86) down to the fainter stars that provide an astrometric and photometric reference, two interleaved exposure times of 6.38 seconds and 2.54 seconds are used, with a camera readout time of 1.55 seconds (see the timing diagram in Figure \ref{fig:timing}).
The long exposure time echoes the observing strategy of MASCARA \citep{snellen12,snellen13} which reaches up to magnitude 4, while the short exposure time ensures that $\beta$ Pictoris is never saturated.
The cameras do not track the sky motion, and so the stars follow fixed declination tracks on the CCDs during the night.
The short integration times minimize the trailing of stellar images on the CCDs, i.e, less than 1.6 pixels for the longer exposure times.
Exposures are synchronised to repeat at the same Local Sidereal Time every night, so that the stars appear on the same pixels at the same LST, aiding the photometric and astrometric calibration.

\subsection{Hardware Design}

\subsubsection{External Enclosure}

The hardware of bRing is housed within an aluminium framework that fits within a volume of $0.75\times 0.75 \times 1.15$ m.
Detachable aluminium panels provide access to the internal hardware, and foam stripping along the inner edges of the panels provide weatherproofing.
All internal components that might require maintenance are accessible through the back door of bRing.
The optics hood is a fully environmentally isolated volume for the cameras and lenses, which prevents dust and moisture contaminating the optics, and is actively temperature controlled.
The base of the enclosure contains a 19-inch computer rack with three blade computers, an electronics control box, a GPS time server, and an uninterruptable power supply.
Rain, snow and dust are minimised from entering the computer enclosure through downward facing vents in the front and back of the main enclosure.
In nominal operation, bRing consumes 1.0 kW of electrical power, with all computers (0.35 kW), cameras and cooling system operating.
An uninterruptable Power Supply rated at 1500VA provides emergency backup for up to 30 minutes for bRing in the event of a short power outage, and the ability to perform a controlled shut down in the case of a longer power outage.

\subsubsection{Temperature Control}

Active cooling is required to prevent overheating of the cameras, computers, and control electronics.
A simplified cooling scheme is currently implemented inside the enclosure.
The optics hood contains the two cameras within a temperature insulated enclosure and is actively temperature controlled with two banks of Peltier devices. 
These Peltier devices are either operated in cooling mode (heat pump) or heating mode (modulating the Peltiers), to allow heating or cooling of the optics hood. 
The heat extracted from the optics hood is passed into the computer enclosure forming the remaining space within bRing.
The computer enclosure is actively cooled with ambient air from the outside of bRing.
An intake vent with fans in the front panel draws in ambient air, which then flows over the computer rack, electronics box, and air is forced through the cooling fins of the Peltier devices with a rack of circulation fans, before passing out through four downward facing extraction fans built into the exhaust vent inside the back door.

\subsubsection{Mechanisms}

One design goal of bRing was to minimise the potential number of mechanical failures by minimising the number of moving parts in the enclosure.
There are two shutters on the optics hood that open and close to provide optical and mechanical protection of the camera windows during the daytime and rain events.
Each shutter is an aluminium lid with a Maxon stepper motor and gear reducer, housed inside the hinge mechanism.
A rubber gasket around the edge of the hinge provides environmental protection for the stepper motor.
Power and control cables pass through holes in the bRing enclosure from the underside of the shutter hings, through to Maxon controllers fastened on the inside of the optics hood.

\subsection{Environmental monitoring}

Four temperature and humidity sensors provide environmental monitoring of the computer enclosure and optics hood at a sampling rate of approximately 1 Hz.
An external weather station sits on a pole elevated 50cm above the top of the bRing enclosure, providing temperature, humidity, rain and sky brightness telemetry to the bRing control computers through a powered USB cable.

\subsubsection*{Control electronics and computers}

The top part of the main enclosure contains an electronics box that controls the Peltiers, fans, and power supplies for all the electronic components in bRing.
Underneath the electronics box sits three off-the-shelf rack-mounted computers.
These are Dell R230 rack servers (Racklive/ASA customized servers for
bRing AU), with a XEON E3-1240v5 (3.5GHz) processor, 32 GB internal memory and fitted with both a 1 TB SSD drive for local processing as well as an 8 TB HDD for longer term storage. 
The two cameras are separately operated by one computer each, and a third computer acts as the control computer that is responsible for all operations.
The control computer contains the control programs that allow autonomous operation of bRing. Generally, human intervention has only been required during unforeseen and extreme events.
The computer decides whether the weather is favourable for observations, opens and closes the shutters, controls the temperature inside the camera enclosure, and sends commands to the computers controlling the cameras.

\begin{figure}
	\centering
	\includegraphics[scale=0.45]{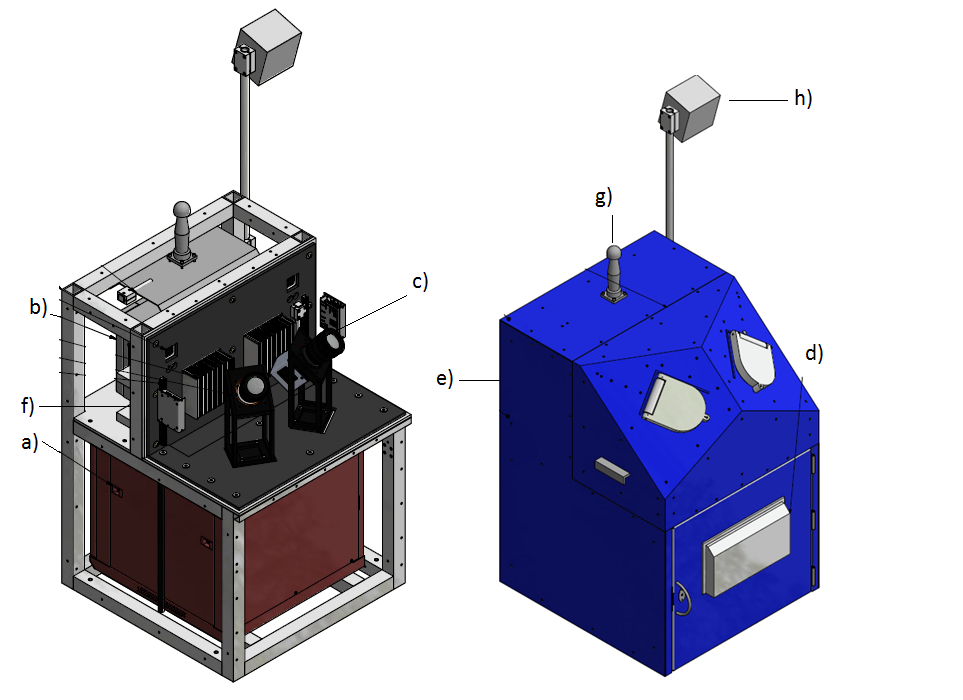}
	\caption{A drawing of the bRing enclosure with and without the cover panels. Labelled are (a) the server rack, (b) the electronics box, and (c) the optics hood. (d) shows the location of the intake fans and (e) the exhaust fans, (f) circulation fans to circulate air in the computer enclosure. The GPS station (g) and the weather station (h) are fixed on the outside of the enclosure.}
	\label{bRing_enclosure_drawing}
\end{figure}

\subsection{Software Design}

\begin{figure*}
	\centering
	\includegraphics[scale=0.55]{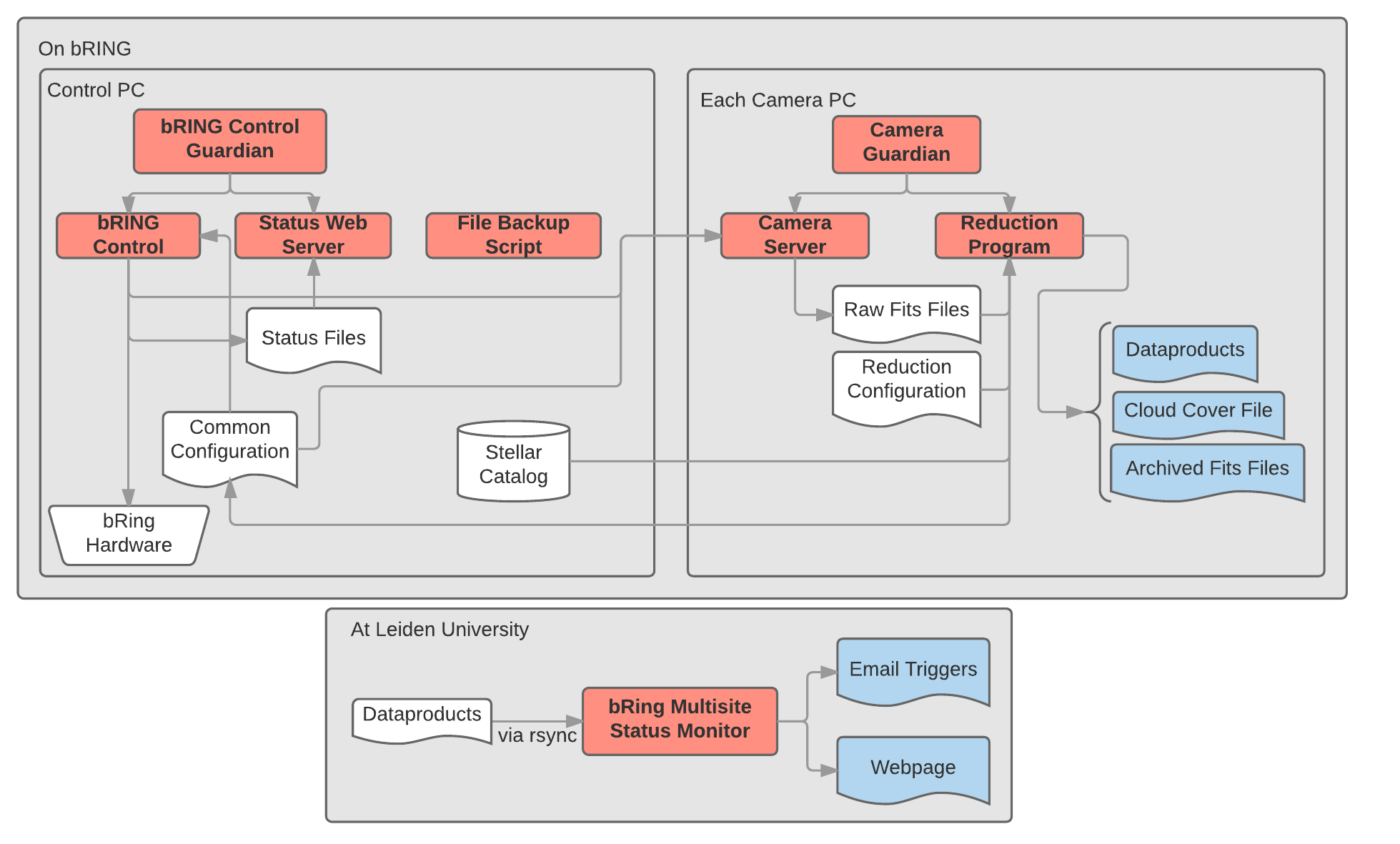}
	\caption{A drawing of the software architecture of the bRing stations.}
	\label{software}
\end{figure*}

The bRing Control Software consists of a small number of custom Python programs along with commodity software running on the three bRing Windows 10 PCs.
Figure \ref{software} shows an overview of the various software components on bRing along with a schematic layout of the systems and programs that exist off-site and are used to support bRing.
One PC is responsible for running the primary station control software, \texttt{bringctrlsrv}, while each of the other computers runs an instance of the camera management software, \texttt{bringcamserv}, and data reduction software, \texttt{bringreduce}. 
These programs are executed and maintained by a guardian script started at boot by the Windows task scheduler.

During normal operations, the \texttt{bringctrlsrv} program monitors station status (e.g. are power supplies and sensors functioning nominally), weather, and the sun altitude to decide if it is both safe and desirable to take data or if it is necessary to notify the bRing team of an anomaly.
It is responsible for maintaining the internal cooling unit, opening and closing the shutters, and notifying the camera software of the current instrument state.
In addition it listens to status messages from the camera PCs to monitor operational status.

The camera software operates as a simple state machine: 1) idle (e.g. daytime), 2) prepare (i.e. time to cool down and take darks), and 3) take exposures matched to the LST index when informed that the shutter is open.
The reduction software is run by the guardian at a lower process priority to ensure there are no timing issues, though in practice we have not found this precaution to be necessary.
Once the night is over, data reduction programs take typically a few hours to reduce the light curves of all stars that are seen by bRing.
The data is then copied over, along with the co-added 50 frame averages, back to Leiden.

\subsection{Sites}

The Sutherland observing station of the South African Astronomical Observatory is located 370 km north east of Cape Town in the Northern Province of South Africa.
It is located 1768m above sea level, where the site is home to over thirteen optical/NIR observatories and has been in operation since 1972.
The median seeing is 1.32 arcseconds and there are no dominant seasonal trends in weather \citep{catala13}.
A strong correlation with wind direction and seeing at the site have shown that poor seeing conditions can be expected with winds from the south-east.
The bRing instrument was installed at the southern end of the plateau, with a clear view of the southern sky down to the horizon.
First light was obtained on 17 January 2017.
The geodetic coordinates of bRing are $-32.3812\pm0.0001$ degrees latitude and $20.8102\pm0.0001$ degrees longitude East at an elevation of 1798m.

The Siding Spring observing station is located at $-31.272189\pm0.0001$
degrees latitude and $149.0622\pm 0.0001$ degrees longitude West at an
elevation of 1165m.

\subsection{On-sky Image Quality}

\begin{figure*}
	\centering
	\includegraphics[scale=0.20]{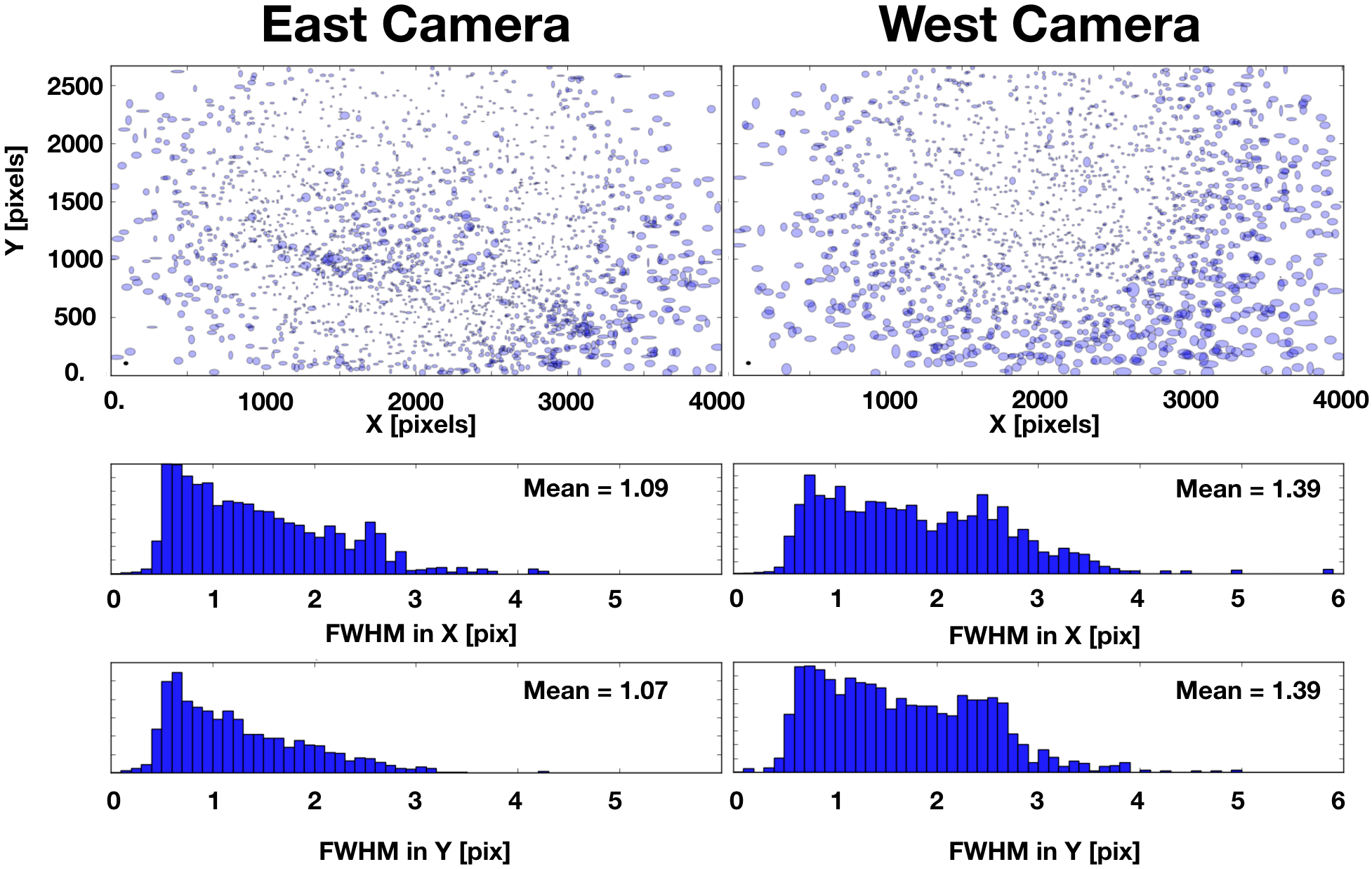}

	\caption{FWHM as a function of location on East and West cameras. The ellipse axes are proportional to the FWHM measured in that axis. The black circle in the lower left corner marks the size for a circular FWHM of 1 pixel.}
	\label{Focus}
\end{figure*}

The nominal focus of the lenses on the CCDs produce a point spread function (PSF) that has a full width half maximum (FWHM) of just less than one pixel for on-axis sources.
The PSF degrades with increasing off-axis angle, with astigmatism and coma increasing significantly towards the edge of the field of view.
The transmission of the optics is a strong function of distance from the optical axis, decreasing to approximately 30\% of the central transmission.
To ensure that the on-axis sources do not saturate the camera, we defocus the lens so that the central PSF has a FWHM of over 2 pixels, and that the line of best focus is in the shape of a ring approximately two thirds of the diameter of the field of view in a region of reduced total transmission.
The focusing procedure was carried out at the bRing site using live acquisition of images and a manual focusing of the lenses.
We fit a Gaussian profile in the $x$ and $y$ axes for every star above a given flux threshold in the East and West cameras, and the measured FWHM are shown in 
Figure \ref{Focus}.

\subsection{Observing Strategy}

%
%
%
To simultaneously obtain good photometry on stars brighter than $m_V=4$ and maximize the amount of data produced for stars in the magnitude range observed by MASCARA, we take CCD images with exposure times of $6.40$ and $2.55$ sidereal seconds, respectively.
Nightly operations with bRing start when the sun altitude passes below $0^\circ$, when a series of dark frames with alternating exposure times are taken for calibration.
As the sun altitude passes below $-10^\circ$, the weather conditions are assessed and if the conditions are suitable for observing the shutters are opened and science observations commence.
Observations are interrupted when the weather conditions are no longer met, and are only resumed after 15 consecutive minutes of good conditions.
When the sun altitude again passes $-10^\circ$ the shutters are closed and a second series of dark frames is taken.

The bRing station produces approximately 120 Gb per night per camera of raw data, which is compressed by 40-60\%.
The stellar light curves and a background region for each star are the only data transferred to the central database in Leiden, Netherlands.
Up to one month of raw images are stored locally and can be retrieved in the case of significant transient events.

For bRing the goal of detecting the Hill sphere transit requires rapid photometric analysis to be performed alongside the data acquisition.
The MASCARA reduction pipeline was modified for this purpose, reducing the incoming data every fifty images.
The $\beta$ Pictoris photometry is processed with a nightly calibration and sent to Leiden every 15 minutes, allowing for both manual and automated detection of photometric excursions.

\subsection{Data Reduction}

The astrometry and photometry are performed on site since the volume of raw data cannot be transported to Leiden at a fast enough rate.
The data reduction is similar to that of MASCARA as described in \citet{Talens17} with a few minor exceptions, necessitated by the alternating exposure time cadence of bRing.
The astrometry is exclusively performed on the long exposures to ensure a sufficient number of stars are available for obtaining a good astrometric solution.
For every fifty frames obtained in long and short cadence, aperture photometry is performed, producing a flux measurement with an associated photometric error.
The reduced photometry obtained from this on-site reduction is transferred to Leiden Observatory where systematic corrections are computed every two weeks using a modified version of the coarse decorrelation algorithm \citep{CollierCameron06,Talens17}. 

The measured instrumental magnitude, $m_{it}$, for star $i$ at time $t$  is corrected using the equation below to generate the reported magnitude $m_{bRing}$ (Talens at al., in prep):

$$ m_{bRing} = m_{it} - c_{qt} - T_{nk} - m_{intrapix} $$

The algorithm iteratively computes corrections for three systematic effects present in the data: the transmission $T_{nk}$ measured at the CCD containing both the pixel quantum efficiency and total throughput of the lenses and sky, intrapixel variations $m_{intrapix}$ caused by the geometry of the cylindrical lenses on the CCD, and temporal variations $c_{qt}$ due to atmospheric transparency and clouds. 

For bRing the corrected data needs to be available within a few hours so that changes in the photometry of $\beta$ Pictoris may be detected and other observing facilities can be notified to perform follow up observations.
In order to facilitate this, the MASCARA systematics removal algorithm was adopted to obtain a preliminary on-site correction every fifty images.
Since the full corrections take several hours to compute, the algorithm was split into a two-step solution process: a complete daytime solution and a partial night time solution.

The daytime solution uses the raw photometry from the last fifteen nights to compute the full corrections for the transmission, intrapixel variations and temporal variations.
The night time solution then uses the transmission and intrapixel corrections from this daytime calibration and computes only a preliminary correction for the temporal variations during the current night.
Keeping the transmission and intrapixel variations fixed in this way is possible, since these effects are stable over longer (week to month) time scales and keeping them fixed means solving is no longer an iterative process, resulting in the speed up necessary to compute preliminary temporal corrections every fifty images.

\begin{figure*}
	\centering
	\includegraphics[scale=0.4]{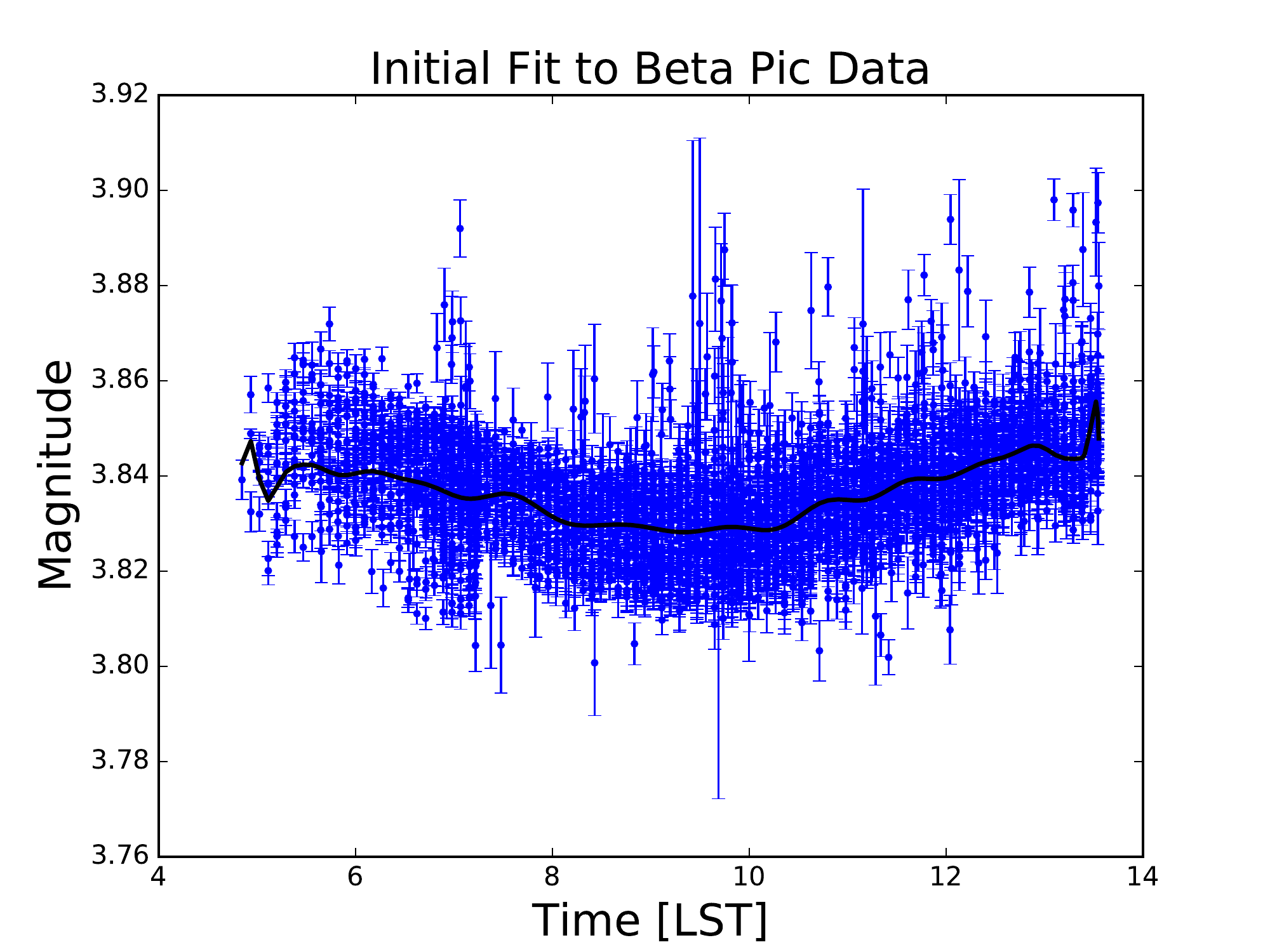}\includegraphics[scale=0.4]{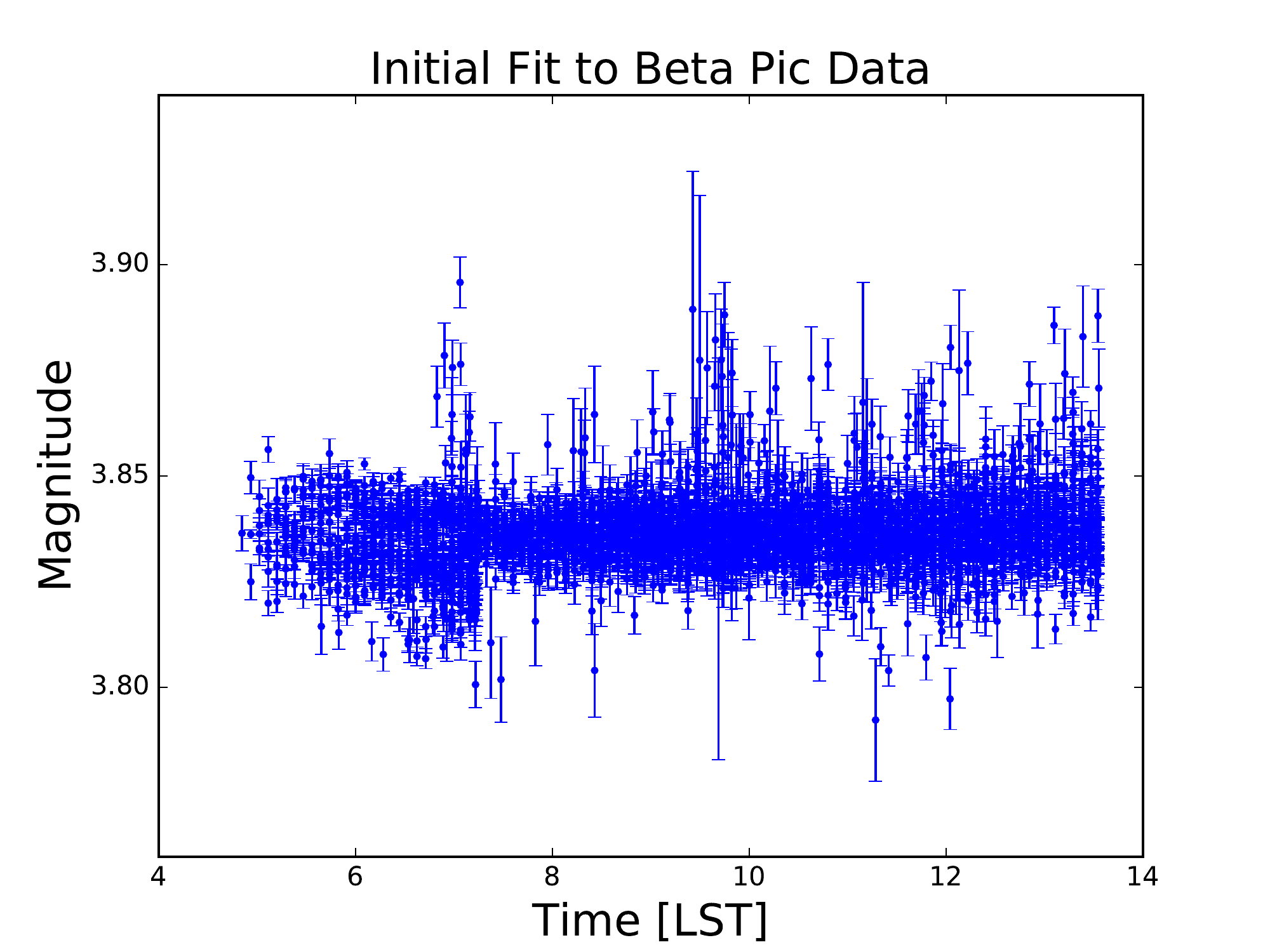}
	\caption{An example data set taken from both cameras, showing the application of the LST correction. The left hand panel is the photometry of $\beta$ Pictoris over two weeks and plotted as a function of LST. The blue values show the original data, and the black values show the calculated correction function. The right hand panel shows the same photometry after the LST flux correction is applied.}
	\label{InitialFit}
\end{figure*}

After these three corrections are applied, we noticed an extra systematic effect on the order of 2\% amplitude that is correlated with the Local Sidereal Time of the observations, which corresponds with the track of the star across the detector.
The PSF of the lenses change with position in the sky and are wavelength dependent.
Stars with different spectral energy distributions will therefore have different PSFs from each other, and that these colour PSFs will change with location on the detector. Furthermore, the bRing has relatively large pixels, and variable contamination from nearby objects is again different for every star. 
This appears as a 2\% systematic change when one star is calibrated with respect to the ensemble mean of all the other stars in the same region of the sky, as seen in Figure \ref{InitialFit}.

We approximate this correction using data from the star itself.
For a single star, a combined fit is made of the long term variations, as represented by Legendre Polynomials in time with typical time scales of $>$ 3 days, and short term variations, represented by Legendre Polynomials in sidereal time with typical time scales of $>0.25$ sidereal hours, using a minimum of two weeks of data.
Two weeks of data are plotted as a function of LST (see left hand panel of Figure \ref{InitialFit}).
The resulting light curve for $\beta$ Pictoris is shown in the right hand panel of Figure \ref{InitialFit}.

\section{First Light and Regular Monitoring of \texorpdfstring{$\beta$}{B} Pictoris}\label{firstlight}

\begin{figure*}
	\centering
	\includegraphics[scale=0.40]{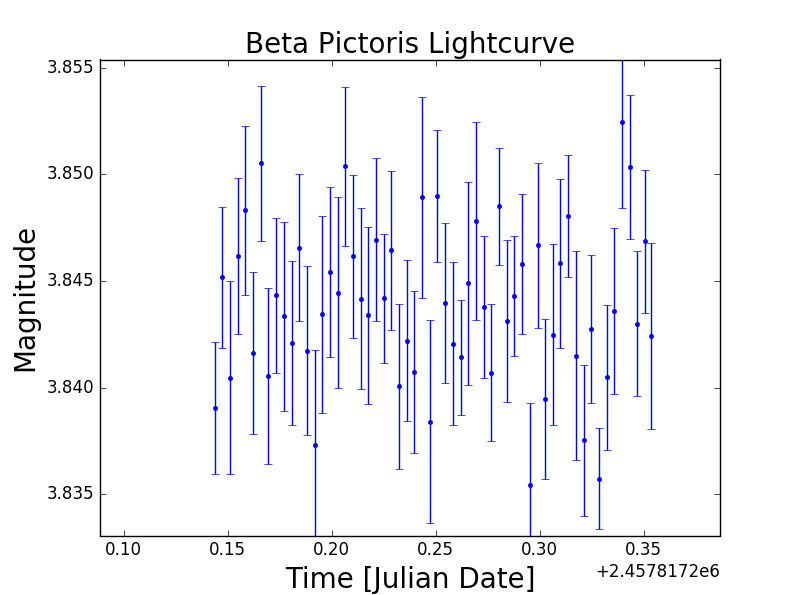}\includegraphics[scale=0.4]{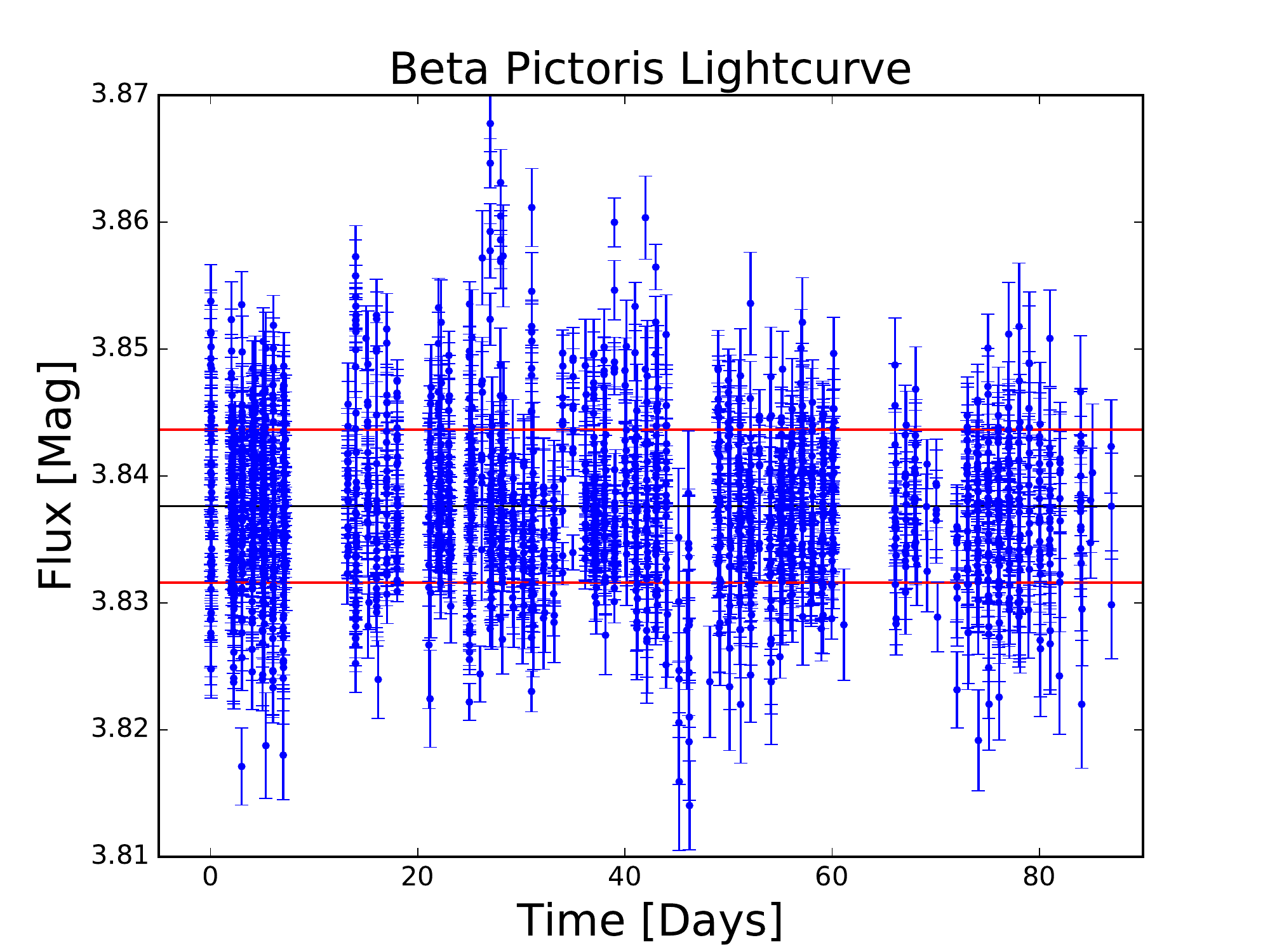}
	\caption{Final bRing photometry of $\beta$ Pictoris for 2017 March 04 (left panel) and for 2017 Feb 01 UT to 2017 March 15. The data is predominantly from the West camera. The black line denotes a mean magnitude of 3.840 magnitudes, and the red lines denote the one-sigma limits of the data.}
	\label{BetaPicLightcurve}
\end{figure*}

First light for bRing was obtained on 17 January 2017.
The bRing observatory has been operating since that time and obtaining observations every clear night.
Due to the location of bRing on the plateau with a clear view of the southern sky down to the horizon, $\beta$ Pictoris will be monitored nightly throughout the year with at least 3 hours of observations per night, even when the sky visibility of $\beta$ Pictoris is the worst in mid-June.
A typical light curve for $\beta$ Pictoris is shown in the left hand panel of Figure \ref{BetaPicLightcurve}, and the first three months of operation is shown in the right hand panel.
The error on the photometry of $\beta$ Pictoris is typically $\pm3$ millimagnitudes every five minutes and the long-term error is $\pm 12$ millimagnitudes, reflecting the presence of the $\delta$ Scuti pulsations in $\beta$ Pictoris.

Triggers for follow up observations are planned, based on changes in the photometry of $\beta$ Pictoris. 
The 1981 photometry showed a rise of 2\% in the flux over a seven day period, followed by a rapid rise of 6\% over 3 days. 
We therefore set our trigger for other observations to be when we see a change in the flux of 2\% for two consecutive days.
Additional verification is provided by other photometric monitoring campaigns of $\beta$ Pictoris, which includes the ASTEP telescope \citep{Crouzet10} at Concordia in Antarctica.
If a significant change in brightness is noted, we confirm by email when possible before sending out a general alert to the astronomical community for spectroscopic monitoring with larger telescopes.

\section{Conclusions}\label{conclusions}

In this paper we describe an observatory dedicated to monitoring the $\beta$ Pictoris b Hill sphere transit from April 2017 through to 2018.
We discuss the design and implementation of the observatory in South Africa, with a second station to follow in Australia to provide continuous longitudinal coverage of the star.
We then describe the data acquisition and reduction and show that we achieve 0.5\% photometric precision every five minutes.
Due to regularly data transfers to Leiden, the photometry is being monitored for photometric dimming or brightening events, which have a characteristic time scale of 4 to 24 hours.  

While bRing is observing $\beta$ Pictoris, a number of other complimentary surveys are also observing for any potential event in $\beta$ Pictoris.  This includes:
\begin{itemize}
\item ASTEP 400, a 40cm telescope located at Concordia, Antarctica with $i'$ band monitoring (PI T. Guillot).
\item AST3-2, one of three 68cm telescopes at Dome A, Antarctica, observing with a $g$ filter (PI L. Wang).
\item Photometric monitoring with the BRITE-Constellation CubeSats (PI K. Zwintz).
\item PICSAT, a CubeSat for dedicated monitoring of $\beta$ Pictoris (PI S. Lacour).
\item Precise high resolution spectroscopic observations with the HARPS instrument at the ESO La Silla 3.6m telescope (PI A.-M. Lagrange).
\item High Resolution Spectrograph observations at the Southern African Large Telescope (PI B. Lomberg).
\item High spectral resolution monitoring with UVES at the Very Large Telescope (PI: E. de Mooij).
\item Hubble Space Telescope monitoring photometrically with WFC3 (PI J. Wang) and spectroscopically with COS (PI P.A. Wilson).
\end{itemize}
Furthermore, a detection of an event in the light curve will trigger high cadence photometry and spectroscopic follow-up with a range of large telescopes.
Beyond 2018, the future of bRing will be to continue photometric observations, with science goals to search for longer period transiting phenomena. 
Over 40,000 stars with magnitudes from 4 to 10 south of a declination of $-25^o$ are being monitored by bRing, providing a large time series database useful for searching for transiting exoplanets and other transient phenomena, and variability studies of stars.

\begin{acknowledgements}

MAK, RS, PD gratefully acknowledge funding from NOVA and Leiden Observatory.
We thank the NWO/NRF for travel funding for this project.
SMC, and BBDL thank NRF PDP.
This work is based on the research supported by the National Research Foundation.
This research made use of Astropy, a community-developed core Python package for Astronomy \citep{2013A&A...558A..33A}.
We thank the SAAO operations/technical staff for their logistical support in the commissioning 
of bRing in Sutherland. 
EEM and SNM acknowledge support from the NASA NExSS program. 
Construction of the bRing observatory to be sited at Siding Springs, Australia would not be possible
without a University of Rochester University Research Award, help from Mike Culver
and Rich Sarkis (UR), and generous donations of time, services, and materials from 
Joe and Debbie Bonvissuto of Freight Expediters, Michael Akkaoui and his team at Tanury Industries, Robert Harris and Michael Fay at BCI, Koch Division, Mark Paup, Dave Mellon, and Ray Miller and the Zippo Tool Room. 
EEM’s contribution to this study was started at the University of Rochester, sponsored by a University Research Award, and was completed at the Jet Propulsion Laboratory, California Institute of Technology, under a contract with the National Aeronautics and Space Administration. Reference herein to any specific commercial product, process, or service by trade name, trademark, manufacturer, or otherwise, does not constitute or imply its endorsement by the United States Government or the Jet Propulsion Laboratory, California Institute of Technology.
This document is approved for unlimited release (CL\#17-4355).
We thank our anonymous referee for comments that improved this manuscript.
\end{acknowledgements}

\bibliographystyle{aa}  
\bibliography{bring}
\end{document}